# FIRST BEAM SPECTRA OF SC THIRD HARMONIC CAVITY AT FLASH*

P. Zhang[†‡#], N. Baboi[#], T. Flisgen[°], H.W. Glock[°], R.M. Jones[†‡],
B. Lorbeer[#], U van Rienen[°], I.R.R. Shinton[†‡]

[†]School of Physics and Astronomy, The University of Manchester, Manchester, U.K.
[‡]The Cockcroft Institute of Accelerator Science and Technology, Daresbury, U.K.
[#]Deutsches Elektronen-Synchrotron (DESY), Hamburg, Germany
[°]Universität Rostock, Rostock, Germany

*Abstract*

Third harmonic superconducting cavities have been designed and fabricated by FNAL to minimise the energy spread along bunches in the FLASH facility at DESY. A module, consisting of four nine-cell 3.9 GHz cavities, has been installed in FLASH. The first measurement with beam excitation is presented, and the comparisons to transmission measurement without beam and simulations are made. Higher order modes (HOMs) are able to propagate to adjacent cavities through attached beam tubes. Modes from 1.3 GHz cavities in the module nearby also propagate into this module.

## INTRODUCTION

FLASH [1] is DESY's free-electron laser user facility providing ultra-short electron bunches with high peak current to generate coherent light with unprecedented brilliance. It is also a user facility for various accelerator studies. Due to the length of the bunch and the sinusoidal RF field of 1.3 GHz, a curvature in the energy-phase plane develops, which leads to long bunch tails and the reduction of peak current in bunch compressions. To linearize or flatten the RF field, harmonics of the fundamental accelerating frequency of the linac are added. In FLASH, third harmonic superconducting cavities operating at 3.9 GHz are used. A cryo-module named ACC39 has been built by FNAL [2] and tested in the Cryo-Module Test Bench (CMTB) facility at DESY [3]. Subsequently it has been installed in FLASH (Fig. 1), after the first accelerating module (ACC1) containing eight 1.3 GHz cavities and before the first bunch compressor. ACC39 is composed of four inter-connected cavities, namely C1 through C4 (illustrated in Fig. 2). The ACC39 module has been recently completed commissioning together with other new components in FLASH [4].

The wakefields in these third harmonic cavities are considerably larger than those in the main accelerating cavities as the iris radius is significantly smaller (15 mm compared to 35 mm in TESLA). Besides that, HOM components of the wakefields are much more complex than those in 1.3 GHz accelerating cavities since larger-diameter beam pipes allow most HOMs to propagate to adjacent cavities. It is important to ensure that the HOMs are well-suppressed and their effect on the beam is minimized by aligning the beam to the electrical centre of the cavity. On the other hand, the beam-excited fields also enable the beam position within the cavity to be remotely determined. For this purpose we plan the design of special electronics.

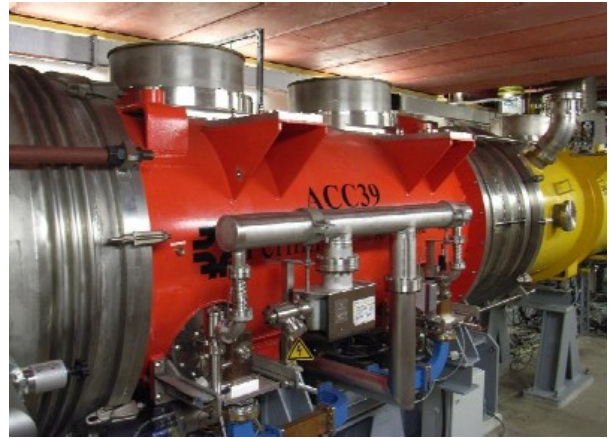

Figure 1: ACC39 module in FLASH beam line. ACC1 can be seen in yellow at the right of the picture.

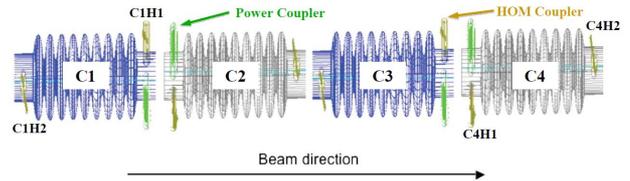

Figure 2: Schematic representation of 4 cavities within ACC39. The power couplers are indicated in green, and the HOM couplers are indicated in brown.

It is vital to have a good understanding of the modal characteristics within the cavities to design electronics. To this end we investigated the spectra of these modes by transmission measurements done both within CMTB facility and after the module installation in FLASH [5]. In this paper we report on the first HOM measurement with beam-excitation on ACC39 recently made at FLASH. Modal spectra are analysed and compared to simulations and to measurements made without beam-excitations.

## BEAM-EXCITED HOM MEASUREMENTS AT FLASH

After ACC39 was connected to HOM board rack outside the FLASH tunnel by cables, a transmission measurement was conducted by using a network analyzer

___

*Work supported by European Commission under the FP7 Research Infrastructures grant agreement No.227579.

(NWA), and this measurement depicted the interactions between HOM couplers and the electromagnetic fields solely induced by the NWA without any beam effects. Afterwards, HOMs were identified and compared with simulations [5]. Then the first beam-excited HOM measurement has been made. This measurement depends on both beam (illustrated in Table 1) and mode properties.

Table 1: Beam properties during the measurement. *BPM denotes Beam Position Monitor (about 4 m away from the center of ACC39), and the mean value and position jitter is calculated by averaging the BPM readings of 66 bunches in each HOM coupler measurement.

| Bunch repetition rate | 10 Hz |
|---|---|
| Beam charge | ~ 500 pC |
| Charge jitter | < 4 % (peak to peak) |
| BPM* reading (Mean) | -0.14 mm (X-offset) |
| | -1.63 mm (Y-offset) |
| Position jitter (RMS) | ~ 50 μm (X-offset) |
| | ~ 30 μm (Y-offset) |
| BPM resolution (RMS) [6] | ~ 20 μm @ 500 pC |

The measurements are conducted using a Tektronix Oscilloscope (Scope) of up to 6 GHz bandwidth and a Tektronix Real-time Spectrum Analyzer (RSA) [7] to take signals from both HOM couplers of all 4 cavities at the ACC39 HOM board rack. A 10 dB external attenuator (another 10 dB internal attenuation is set in RSA for spectra from 3.8 GHz to 4.0 GHz) is connected to each HOM coupler to reduce the power of beam-excited signal. Time-domain signals and real-time spectra were recorded.

The time-domain HOM signal is sampled in 20 GS/s and 8000 points with 400 ns time length for each measurement. Typical recorded waveforms from both HOM couplers of C1 are shown in Fig. 3. Each waveform is excited by a single electron bunch, by triggering the Scope with a 10 Hz pulse synchronous trigger. The same trigger is used for RSA measurement.

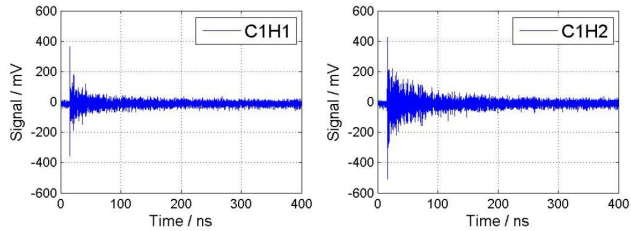

Figure 3: HOM signal from two HOM couplers of C1.

The HOM spectra are recorded over a frequency range of 3.5 GHz to 10 GHz with a frequency interval of 100 MHz and a frequency step of 10 kHz, together with a resolution bandwidth of 100 kHz. The spectrum of each 100 MHz frequency interval was excited by a single bunch. Therefore 65 spectra (100 MHz each) excited by 65 electron bunches (single bunch for each 100 MHz spectrum) are recorded for each HOM coupler. One needs to note that the beam jitter is comparable to the resolution of the BPM reading. Both beam position and charge jitter only affect the amplitudes of the beam-excited HOMs without changing their frequencies.

The first monopole band excited by the beam is consistent with previous transmission measurement without beam-excitation conducted after the installation of ACC39 in FLASH [5], namely NWA (illustrated in Fig. 4). The results of MAFIA [9] simulations are marked by green lines [2]. The differences to the simulations are due to the non-ideal shape of the cavities.

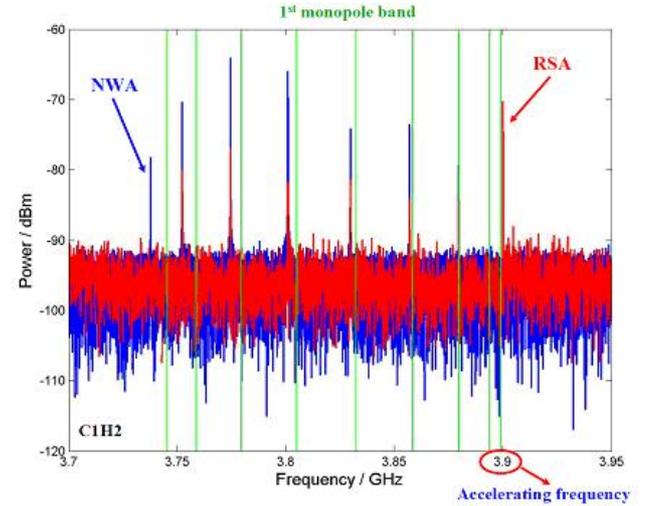

Figure 4: Comparison of spectrum taken by RSA (concatenation of two spectra) and NWA (transmission measurement without beam-excitation) of C1H2. The vertical lines indicate the simulation results.

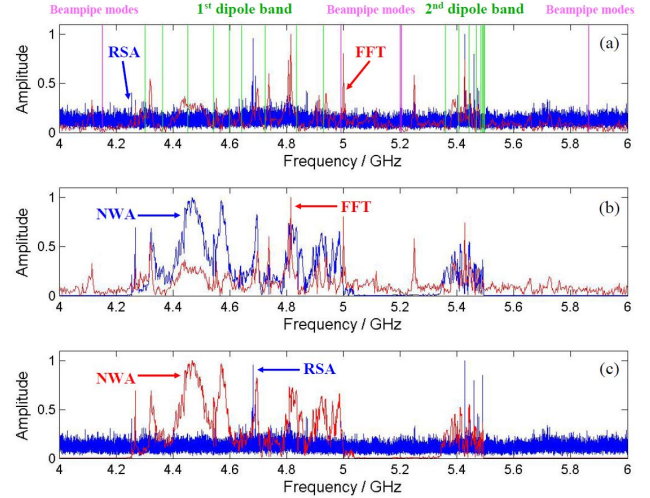

Figure 5: Comparison of spectra taken by Scope (after FFT), RSA (after concatenation) of C1H1 and NWA (transmission measurement without beam-excitation). The vertical lines indicate the simulation results. All spectra are normalized to the highest amplitude in frequency span of 4 GHz to 6 GHz.

In Fig. 5 we show the comparison of spectra (in pairs in each figure) taken from coupler C1H1 (illustrated in Fig. 2) by various devices covering dipole modes between

4 and 6 GHz. A Fast Fourier Transform (FFT) is applied to Scope data, from which the frequency step is calculated as 2.5 MHz. It clearly broadens the peaks in FFT spectrum (typical Q of these modes is $10^4 \sim 10^5$). The RSA spectrum is generated by concatenating 20 spectra of 100 MHz frequency intervals.

Comparisons of spectra taken from two HOM couplers of C1 are presented in Fig. 6. Both Scope and RSA signals are shown. Due to different orientations of HOM couplers installed on both ends of each cavity, the couplers have different response to the beam-excited HOM signal. It is noted that C1H2 appears to have more high-amplitude peaks in frequency region above 5.7 GHz in spectra from both devices due to some modes traveling from ACC1 (illustrated in Fig. 7).

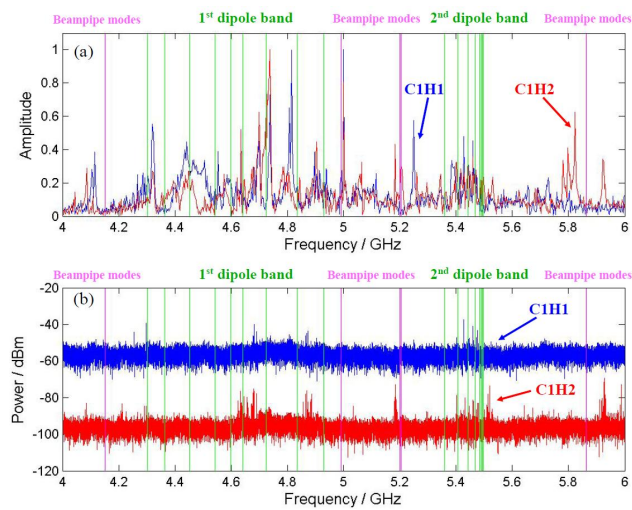

Figure 6: Comparison of spectra taken from C1H1 and C1H2 by using Scope (a) (after FFT) and RSA (b) (after concatenation). The vertical lines indicate the simulation results. (a) is normalized to the highest amplitude in frequency span of 4 GHz to 6 GHz; (b) 40 dBm are added arbitrarily to C1H1 spectrum only for better comparisons.

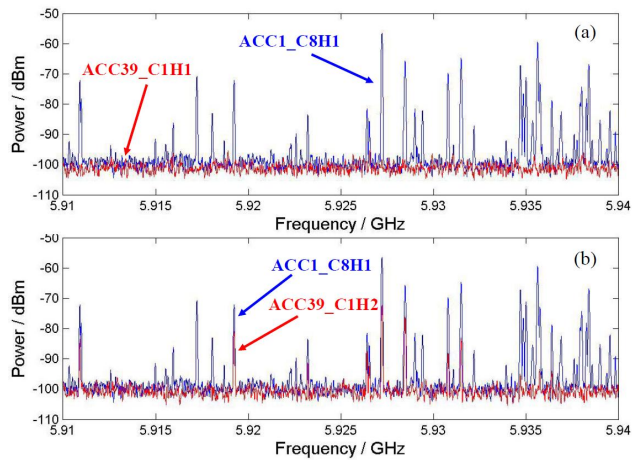

Figure 7: Transmissions from ACC1 to ACC39. Both C1H1 (a) and C1H2 (b) are compared to the last HOM coupler of last cavity (C8H1) of ACC1.

In Fig. 8, we compare the first quadrupole band [8] from RSA and NWA for both HOM couplers of C1. Some quadrupole modes are excited by the beam.

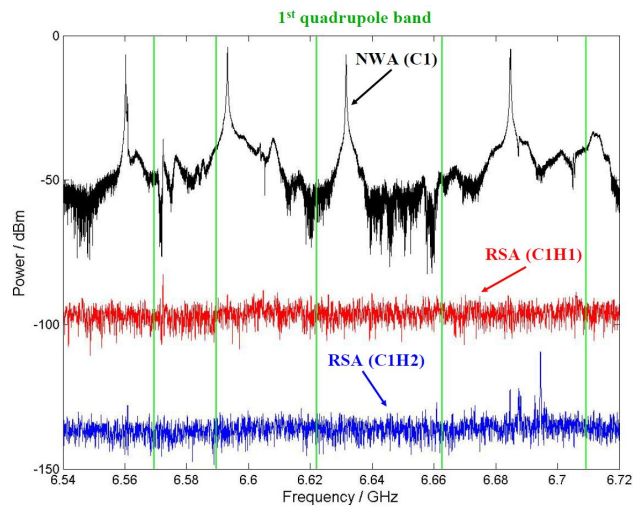

Figure 8: Comparison of spectrum taken by RSA (concatenation of two bunches) and NWA (transmission measurement) of C1H1 and C1H2. The vertical lines indicate the simulation results. 45 dB is added arbitrarily to NWA spectrum while 40 dBm is subtracted from C1H2 spectrum only for better comparisons.

## CONCLUSIONS

Third harmonic cavities have been installed at FLASH in order to linearize the RF field and hence improve the peak current of the linac. The goal of our work is to instrument the cavities with beam diagnostics. To this end, we have made an initial investigation into the modal spectra of the cavities. The beam-excited spectra have been compared to transmission measurements previously made at FLASH. Future work will be concerned with an additional characterization of the higher order modes both with and without beam-excitation, detailed analysis of HOM transmission from the first accelerating module (ACC1) to the third harmonic cavity module (ACC39), in order to determine a mode or group of modes suitable for the electronics for beam and cavity alignment monitoring.